\documentclass{article}
\usepackage{graphicx}
\usepackage{graphics}
\usepackage{amsfonts}
\usepackage{marvosym}

\begin{document}
\title{Current Trends in Evolving Specialization in 
UK Universities}

\author{Fionn Murtagh (1, 2) \\
(1) Science Foundation Ireland, Wilton Place, Dublin 2, Ireland \\
(2) Department of Computer Science, Royal Holloway, \\
University of London, Egham TW20 0EX, England \\
Email: fmurtagh@acm.org}

\maketitle

\begin{abstract}
There are very significant changes taking place in the university 
sector and in 
related higher education institutes in many parts of the world.  
In this work we look at financial data from 2010 and 2011 
from the UK higher education sector.  
Situating ourselves to begin with in the context of 
teaching versus research in universities, we look at the data 
in order to explore the new divergence between the broad agendas
of teaching and research in universities.   The innovation 
agenda has become at least equal to the research and teaching 
objectives of universities.  
From the financial data, published in the Times Higher
Education weekly newspaper, we explore the 
interesting contrast, and very opposite orientations, in 
specialization of universities in the UK.  We find a polarity in 
specialism that goes considerably beyond the usual one of research-led  
elite versus more teaching-oriented new universities.  
Instead we point to the role of medical/bioscience research income 
in the former, and economic and business sectoral niche player roles 
in the latter.  
\end{abstract}

\noindent
{\bf Keywords:}  research funding, student recruitment, 
budget, costs, higher education, university, finance, economics, United
Kingdom, correspondence analysis, multivariate data analysis.

\section{Introduction}

Jenkins (2004) cites Barnett (2003, p. 157): ``the twentieth century saw 
the university change from a site in which teaching and research stood in 
a reasonably comfortable relationship with each other to one in which they 
became mutually antagonistic''.    In his Conclusions, Jenkins (2004, p. 31) 
states: ``From the UK and the USA there is clear evidence that national policies 
and funding for research … has resulted in structural separations between research 
and teaching within the institution.''  

The traditional view of the university encompassing teaching and research is changing.   
Jenkins (2004, p. 5) had earlier noted the following: ``... we may have to move away 
from seeing or disputing a single teaching-research nexus, and develop our 
understanding of the diverse and heterogeneous ways in which teaching and 
research are linked or not.'' Also (p. 31): ``It is possible for institutions with different 
resources and missions to shape and deliver a view of the teaching-research 
nexus that reflects the resources available.''  

Jenkins (2005, p. 9), in discussing ``teaching `only' and research-intensive institutions'' 
concludes (p. 50) that creation of `teaching-only' universities is not justified; also that 
the aspiration should be that ``all students in all higher education institutions learn in a 
research environment''.  

To begin with, therefore, we note this change whereby fairly complete harmony (assuming that was 
once the aspiration or perception) between the teaching and research agendas is no more.   
In this article, we aim to look at data in order to explore this new divergence between teaching and research in universities.   
An econometric model focused on the trade-off by universities of teaching and research is 
pursued by Beath et al. (2011).   Our methodology in this work owes more to Benz\'ecri and 
perhaps Bourdieu too, in that we want to let the data reveal itself in the first instance, 
and then, following on from that, model the data.  See e.g. Lebaron (2011).  

In addition to the teaching and research agendas, it is our view that the business agenda has 
come to the fore in recent times, increasingly on a par with teaching and 
with research being ever more closely aligned with this business agenda. 
  
The plan can be viewed in the following terms, with the emphasis in the 
original (BIS, 2010): 
``Research Councils and Funding Councils will be able to focus their contribution on 
{\em promoting impact through excellent research, supporting the growth agenda.}  They will provide 
strong incentives and rewards for universities to improve further their relationships with business 
and deliver even more impact in relation to the economy and society.'' 
  
Hence we have the new orientation in the university arena that has become 
prominent in recent years, and has been strongly propelled 
forward by the economic downturn following the great banking and (in some countries) real estate 
crash of 2008.     The objective is ever increasingly  becoming: ``to foster more effective 
collaboration between universities and business in the years ahead'' (McMillan 
et al., 2010, p.\ 3).  

That the business innovation agenda has become central to the higher education sector is not in 
doubt.    It is our implicit viewpoint in this work that innovation, understood as encompassing 
business, entrepreneurial and economic activities, has come to be on a par with research and 
teaching.    However our article is not dependent on accepting this viewpoint, either entirely 
or in terms of the change being deep -- although we ourselves take this view.   

One motivation for the turn towards innovation in this sense is additional earning potential 
through ``third stream'' income .  Hatakenaka (2005) considers this in the UK university context.     
Apart from such third stream income, there is also human capital being more aligned with business needs.   
Having innovation on a par with teaching and research as a new characteristic of higher education is 
the main motivation for this article.   Within this context, it seems clear that the institution of 
university is changing.   Some examples of such influence include entrepreneurial course modules or 
other forms of business oriented activity on Masters, or undergraduate courses, and on structured 
graduate training that is part of PhD programmes.    
We can note also the debate around the future role of the PhD degree (see 
the journal {\em Nature}, volume 472, issue of 21 April 2011).

Reasoning further, if the economic crisis post-2008 is engendering change in the higher education 
system, as elsewhere, through government and other agencies being strapped for funds to dispense, 
for research, teaching and associated business growth, then what does empirical data have to reveal 
in regard to this?    

This article is based on two revealing data sets that provide a snap shot of the financial health of 
UK universities.   We looked at this data in order to see what sort of institutional approaches 
seem to be doing well, in the contemporary economical climate.

\subsection{Data and Objectives}

Brief background on the UK universities and funding system can be found 
at, for example, 
http://www.internationalstaff.ac.uk/universities\_in\_the\_uk.php

The data on financial health and safety of UK universities and 
other similar third level or higher education institutes that were 
published in the Times Higher Education newspaper have their own 
tale to tell, as we will show in this article.  We use in particular
the reports of Newman (2010) and Baker (2011).  In Newman (2010) it was 
noted how the downturn in Government spending led to university budgetary 
deficit in many cases, but this was coupled with strong student 
demand and with cost inflation being very much down.  One year later, 
Baker (2011) pointed to overall university finances being ``fairly
healthy'' but pointed to approaching turbulent times, affecting State
funding of the sector, and also undergraduate, postgraduate and 
non-European (incurring higher registration fees) recruitment.  
Baker (2011) referred to the ``oncoming tempest'', of a sustainability 
sort.  

All aspects of the sector's financials are relevant here, 
including salaries, pensions and pension commitments, as well as student
recruitment.  Research funding is important too, although there have been
major changes in regard to this in recent years: ``Even research income --
protected by the government overall -- is a problem for the majority 
owing to the increasing concentration of funding on a small band of 
institutions.''  

Against this backdrop, and based on university financials, we 
ask what has been the higher education system's response.   To 
avoid indebtedness, what sort of role or profile is the university adopting?  

Our methodology is based on ``letting the data speak'', to begin with, 
followed by drilling down in the data, in pursuit of patterns or trends.  
In the case of a need to test hypotheses statistically, the most 
straightforward approach to take is then based on randomization tests.  
Our goal here though is to discern clearcut patterns or trends, and to 
show the current state of play in regard to relative 
positioning of universities.   

By studying such ``relative positioning'' we seek to inform and 
influence policy and decision making.   
In Murtagh (2010) resulting from the Sixth Annual Boole Lecture 
(organized by the Boole Centre for Research in Informatics, 
http://www.bcri.ucc.ie) in 2008, 
we show how information focusing is carried out in data analysis, i.e.\ 
determining where the data is put under the analytic microscope. One 
issue addressed is coverage and completeness of research funding in 
technological sector domains.  Another issue addressed is evolution 
of funding decisions over time. 
We show how the narrative of science and 
engineering policy -- the story that policy decisions have to tell -- 
can be mapped out from the raw data. The orientation of such narrative 
is crucial.  

In this present work, we use the same data analysis approach, 
Correspondence Analysis. 
Based on the data on UK HEI (higher education institute) financials 
provided by Newman (2010), we look for underlying patterns of 
particular interest.  The data is due to accountancy firm 
Grant Thornton and is based on institutions' financial statements
for 2008-2009.   In section \ref{newsect5} we look at data from
2009-2010.  

In Murtagh (2010) we provide background on the  analysis approach 
which takes cross-tabulations as inputs -- in this case of HEIs crossed
with financials on a set of incomes or expenditures.  {\em Profiles}
of the (positively-valued) data, on either rows (i.e., HEIs) 
or columns (i.e.\ financial incomes or expenditures) are mapped
into the same visualizable (hence Euclidean distance-based) space.  
{\em Profiles} are values in the row or column that are divided
by the row/column total.  Hence HEIs, or financial attributes, 
are normalized in this way -- by dividing by their respective 
row/column totals.  

A range of analysis options are  opened up by the Correspondence 
Analysis: simultaneous
display of HEIs and incomes/expenditures; optimal planar display;
accounting for most of the information content (in a precise
mathematical sense) of the data; among others.  

\section{Attributes and Interpretation of the Planar Visualization}
\label{sect2}

Attributes used in the main analysis were as follows.  These 
attributes constituted the primary data used on the 155 institutions.  

\begin{itemize}
\item Attribute 3, Funding council grants
(all grants of: HEFCE, Higher Education Funding Council for England;  or HEFCW, 
Higher Education Funding Council for Wales; or  SFC, Scottish Funding 
Council).
\item Attribute 4, Research grants and contracts
(from all sources other than HEFCE/HEFCW/SFC).
\item Attribute 5, Tuition fees and education contracts (excluding overseas,
i.e.\ non-European resident).
(UK and European including short courses or other ancillary teaching).
\item Attribute 6, Overseas fees.
\item Attribute 7, Other income (from catering, residential, possibly from 
companies spun out).
\item Attribute 8, Endowment and investment income.
\item Attribute 11, Total staff costs (including social security and pension 
contributions).
\item Attribute 13, Total borrowing.
\end{itemize}

We omitted net surplus (attribute 1) because of the remarks in 
the Times Higher, noting how Cambridge had the largest deficit but 
it was a very small percentage of its total income; Bucks New University
recorded a large deficit but then sold a campus to reverse this deficit; 
and Thames Valley University had a surplus but this disappeared when 
HEFCE was reimbursed  for this university's 
over-reporting of its fundable student numbers. Our interest lies in
financial health.  Arising from this, 
we were interested not in the financial position as such  
but rather in determining underlying indications of where the 
sector is headed as it seeks to address the current economic climate.  
So we used the more basic financial data.

Attributes projected into the analysis subsequently were as follows.
These were attributes derived from the more basic data.  

\begin{itemize}
\item Attribute 2, Net surplus as \% of income.
\item Attribute 10, Funding council grants as \% of income.
\item Attribute 12, Total staff costs as \% of income.
\item Attribute 14, Total borrowings as \% of income.
\end{itemize}

\begin{figure}
\includegraphics[width=12cm]{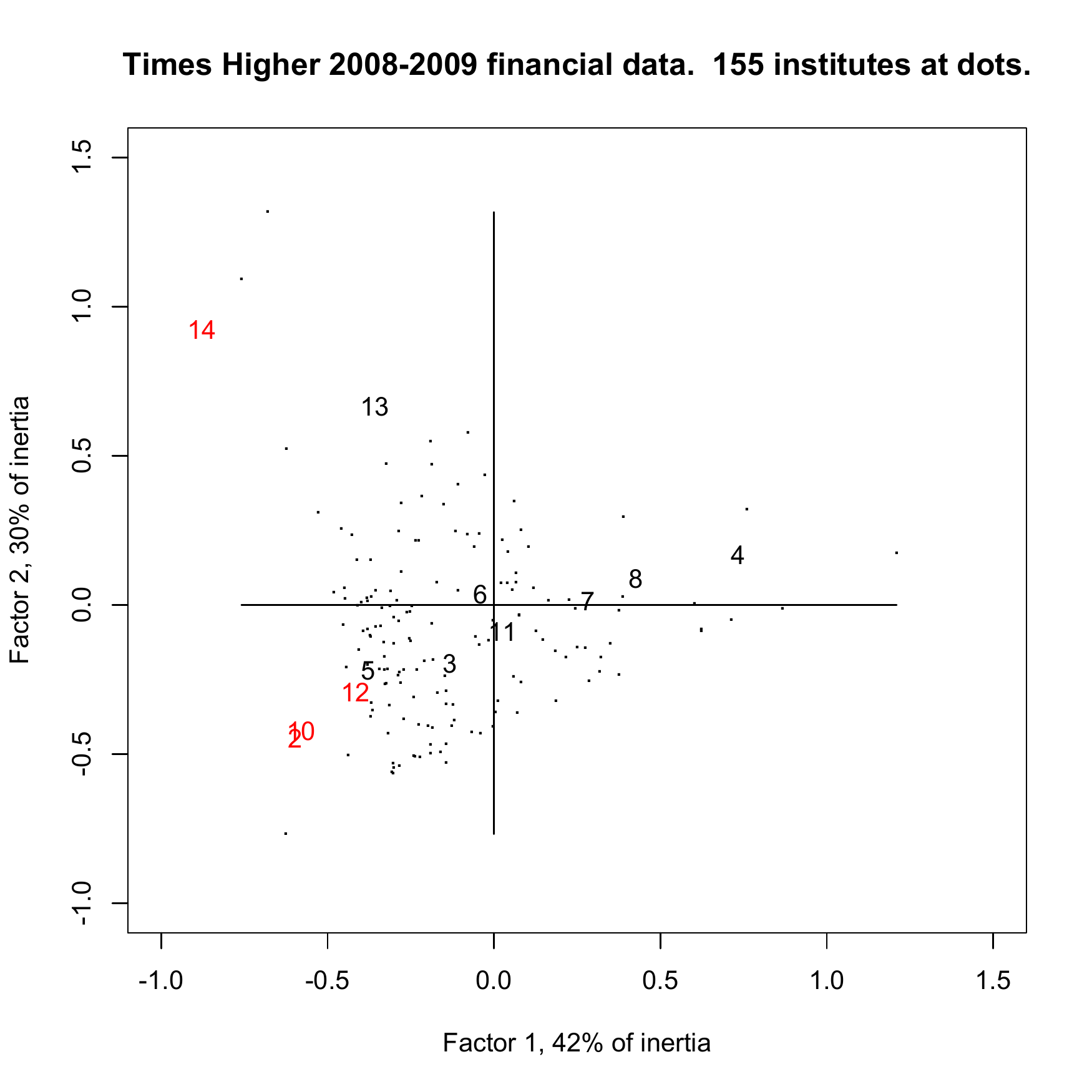}
\caption{A first visualization of the data: the main analysis 
is based on the data table for 155 UK higher education institutions  
using the Times Higher attributes (see text for these) 
3, 4, 5, 6, 7, 8, 11, and 13.  Their locations can be seen in the
(numeric values in the) 
planar projection.  In addition, based on the analysis of the 
main data, the locations were found for the more ``illustrative'' 
attributes, 2, 10, 12, and 14.  These latter are shown in red.  
The higher education institutes, in order not to crowd this 
initial display, are each shown as a dot.}
\label{fig1}
\end{figure}

Fig.\ \ref{fig1} summarizes the data.  Shown in the figure is a 
principal plane projection, accounting for 42 + 30 = 72 \% of the 
information content -- most, therefore.  

Factor 1 is dominated in influence by attribute 4, ``Research 
grants and contracts''.  Such domination is determined not 
just by its relatively extreme (positive or negative) projection 
on this first (newly determined) coordinate axis, but also by
its {\em contribution} to, and its {\em correlation} with, the
first axis.  (Contribution, correlation, inertia expressing 
information, factor, and so on are all mathematically defined 
terms in the Correspondence Analysis data analysis and display 
context.)  

For Factor 2, the dominant attribute is 13, ``Total borrowing'', 
and attribute 11, ``Staff costs'', is not far behind in terms of 
influence.  

On Factor 2, it can be seen that attribute 14, ``Total borrowing 
as \% of income'', is in the same general region as 13, ``Total 
borrowing''.  We can note too that, counterposed to 13 and 14, there 
are attributes 2 (surplus-related), 10 (funding council grants-related),
12 (staff costs-related), and also 5 (tuition fees) and 3 (funding 
council grants) -- all possible countervailing means relative to 
borrowing.  

Interestingly, 
attributes 2 and 10 -- ``Net surplus as \% of income''; and 
``Funding council grants as \% of total income'' -- are closely
located, indicating that the information conveyed is very similar.
Attribute 6, ``Overseas fees'', is close to the origin of the 
display, indicating where it is a not very discriminating attribute
here.

\begin{figure}
\includegraphics[width=12cm]{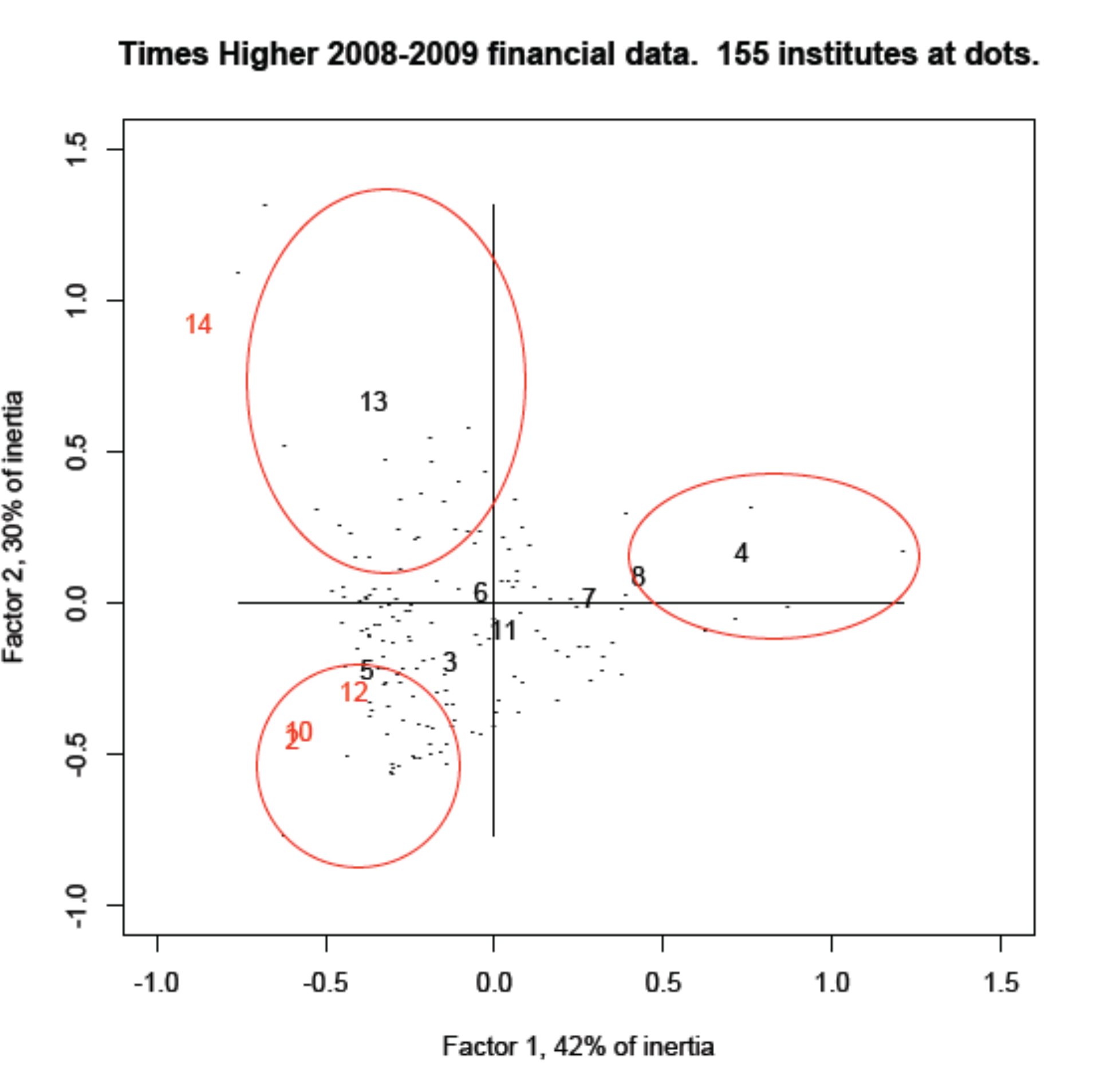}
\caption{We will focus attention on the rightmost HEIs here; on the upper
left ones; and finally on the lower rightmost.  Meanwhile
both HEIs and attributes that are close to the origin (coordinate
0, 0) are average, relating either to average HEI profile, or 
to average attribute profile.}
\label{fig1annot}
\end{figure}

Fig.\ \ref{fig1annot} is the same as Fig.\ \ref{fig1}, just
showing the areas where we will now mostly focus our attention.  

\section{Factor 1: Role of Medical Disciplines in HEIs that are 
Strong in Research Funding}

\begin{figure}
\includegraphics[width=12cm]{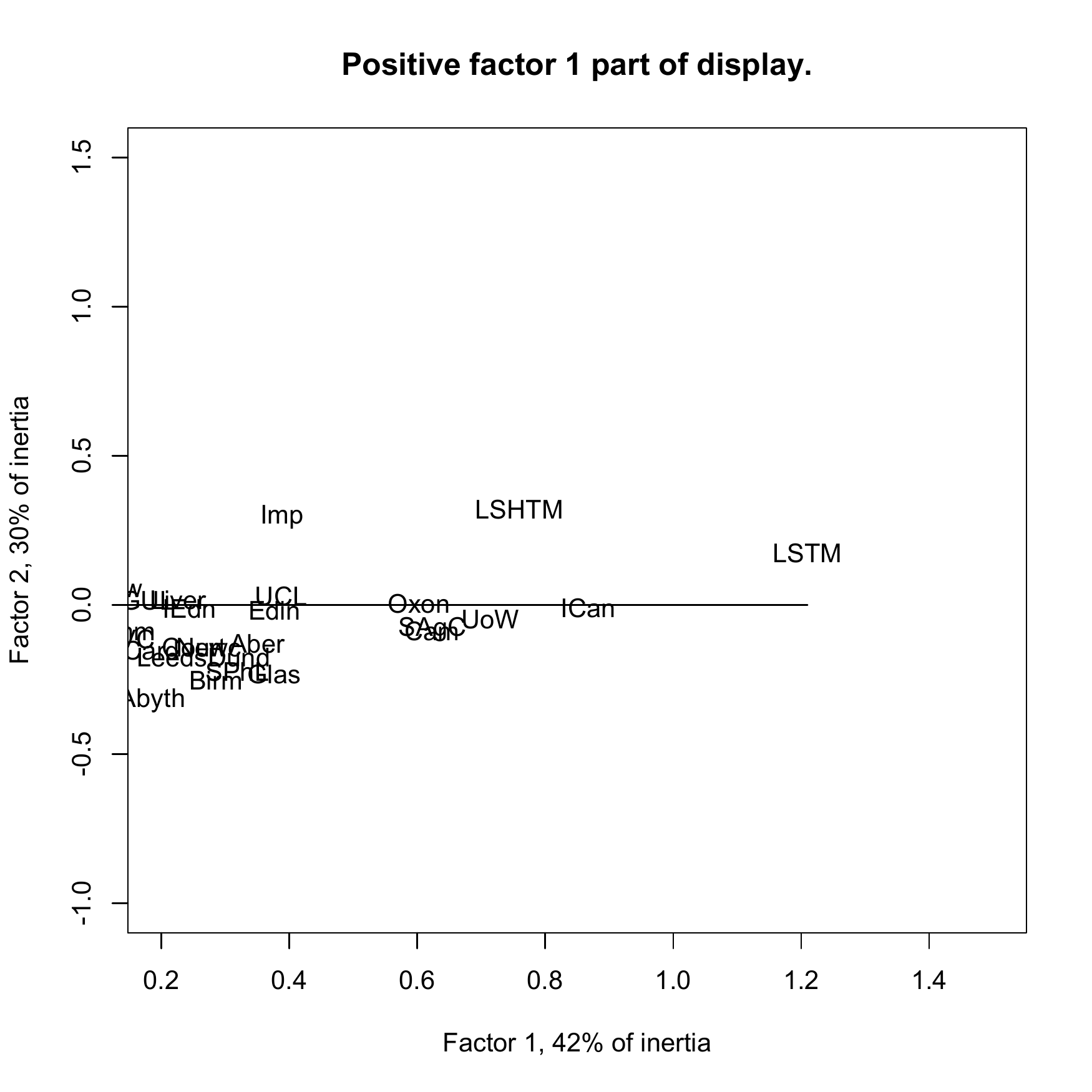}
\caption{The higher education institutes: first, the rightmost 
part of Fig.\ \ref{fig1}, relating to the positive end of Factor 1.}
\label{fig2}
\end{figure}

The most positively linked institutes relative to Factor 1
are to be seen in Fig.\ \ref{fig2}.  These are: 

\begin{itemize}
\item Cambridge (``Cam'', overlapping ``SAgC'')
\item Institute of Cancer Research (``ICan'')
\item Liverpool Sch Tropical Medicine (``LTSM'')
\item Tropical Medicine -- London Sch Hygiene \& (``LSHTM'')
\item Oxford (``Oxon'')
\item Scottish Agricultural College (``SAgC'', overlaid on ``Cam'')
\item University of Wales (an administration only institute) (``UoW'')
\end{itemize}

Somewhat less pronounced in terms of this factor are: Imperial,
UCL (University College London) and University of Edinburgh.  

Apart from the traditionally strong Oxbridge research presence, what 
is also noteworthy is the medical and biosciences presence, albeit 
specialist, in this cluster.  

Adams and Gurney (2010) point to how citation ratings
from Thomson Reuters attribute the lion's share of UK research outcomes 
to five HEIs: Oxford, Cambridge, Imperial, UCL, and LSE (London School 
of Economics).  In our concluding section below we will return to this
view of performance and achievement evaluation.  

\section{Factor 2: Borrowing}
\label{sect4}

\begin{figure}
\includegraphics[width=12cm]{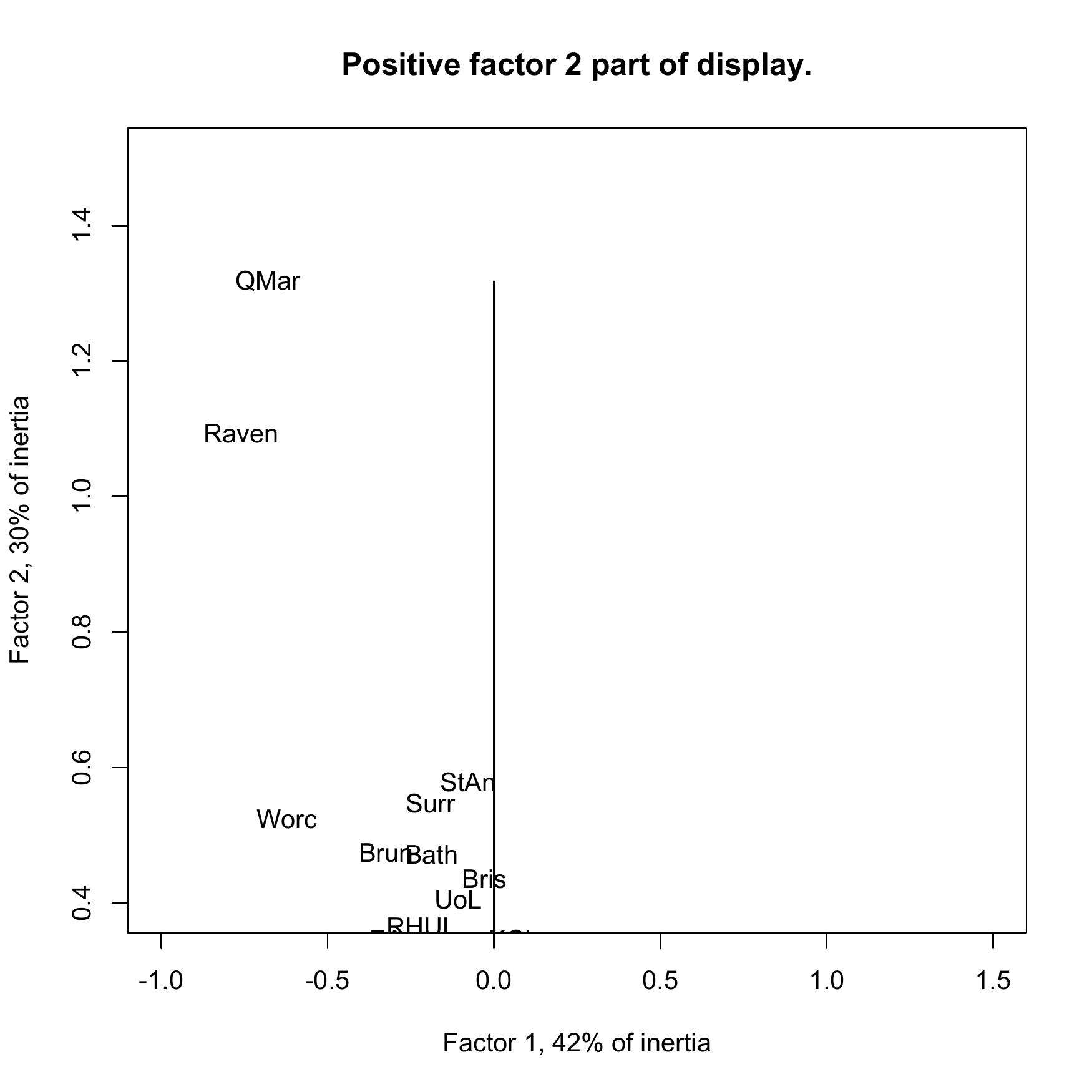}
\caption{The higher education institutes: the positive end of 
Factor 2, cf.\ the complete view in Fig.\ \ref{fig1}.}
\label{fig3}
\end{figure}

As noted Factor 2 is firstly and foremostly related to borrowing.
Fig.\ \ref{fig3} shows the positive end of this factor.  We see 
a number of institutions that are flagged in the Times Higher 
article in terms of high gearing, i.e.\ ``Total borrowings as \& of 
income'':  Queen Margaret University, 
220.5\% of income; Ravensbourne College, 171\%; University of Worcester,
82.5\%; University of Surrey, 63\%; and Brunel University and the 
University of St Andrews, both 62\%.   

\begin{figure}
\includegraphics[width=12cm]{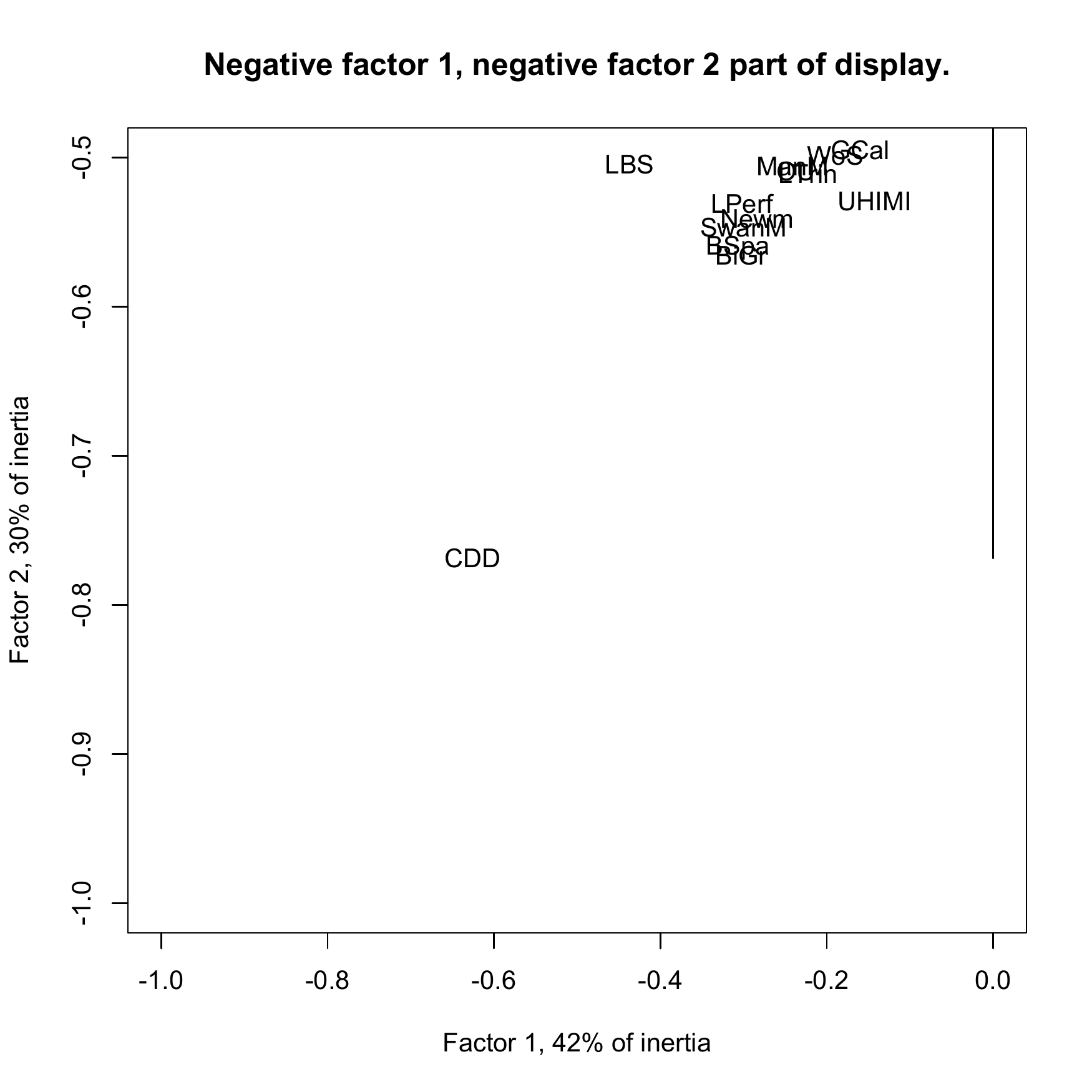}
\caption{The higher education institutes: the negative ends of 
Factor 2 and of Factor 1, cf.\ the complete view in Fig.\ \ref{fig1}.
These are less research funding-based, and also non-borrowings geared, 
institutions.}
\label{fig4}
\end{figure}

We will next look at the non-geared end of Factor 2.  We look at
what is most opposite the research, Oxbridge, medical and biosciences,
end of Factor 1.  What we find in Fig.\ \ref{fig4} is that the 
following institutes are to be found there:

\begin{itemize}
\item Conservatoire Dance \& Drama (``CDD'')
\item Bishop Grosseteste (``BiGr'')
\item Bath Spa (``BSpa'')
\item Swansea Metropolitan (``SwanM'')
\item Newman College (``Newm'')
\item Liverpool Inst Performing Arts (``LPerf'')
\item UHI Millennium Institute (``UHIMI'')
\item Leeds Trinity (``LTrin'')
\item Manchester Metropolitan (``ManM'')
\item Open University (``OU'')
\item London Business School (``LBS'')
\item West of Scotland (``WoS'')
\item Glasgow Caledonian (``GCal'')
\end{itemize}

We note specialist and/or business -- or business sector -- 
orientations that are well represented among these institutions.  
Note again that these institutions are not at all as highly 
geared as those institutions that are more towards the positive 
end of Factor 2.  

\section{From 2008-2009 to 2009-2010}
\label{newsect5}

\begin{figure}
\includegraphics[width=12cm]{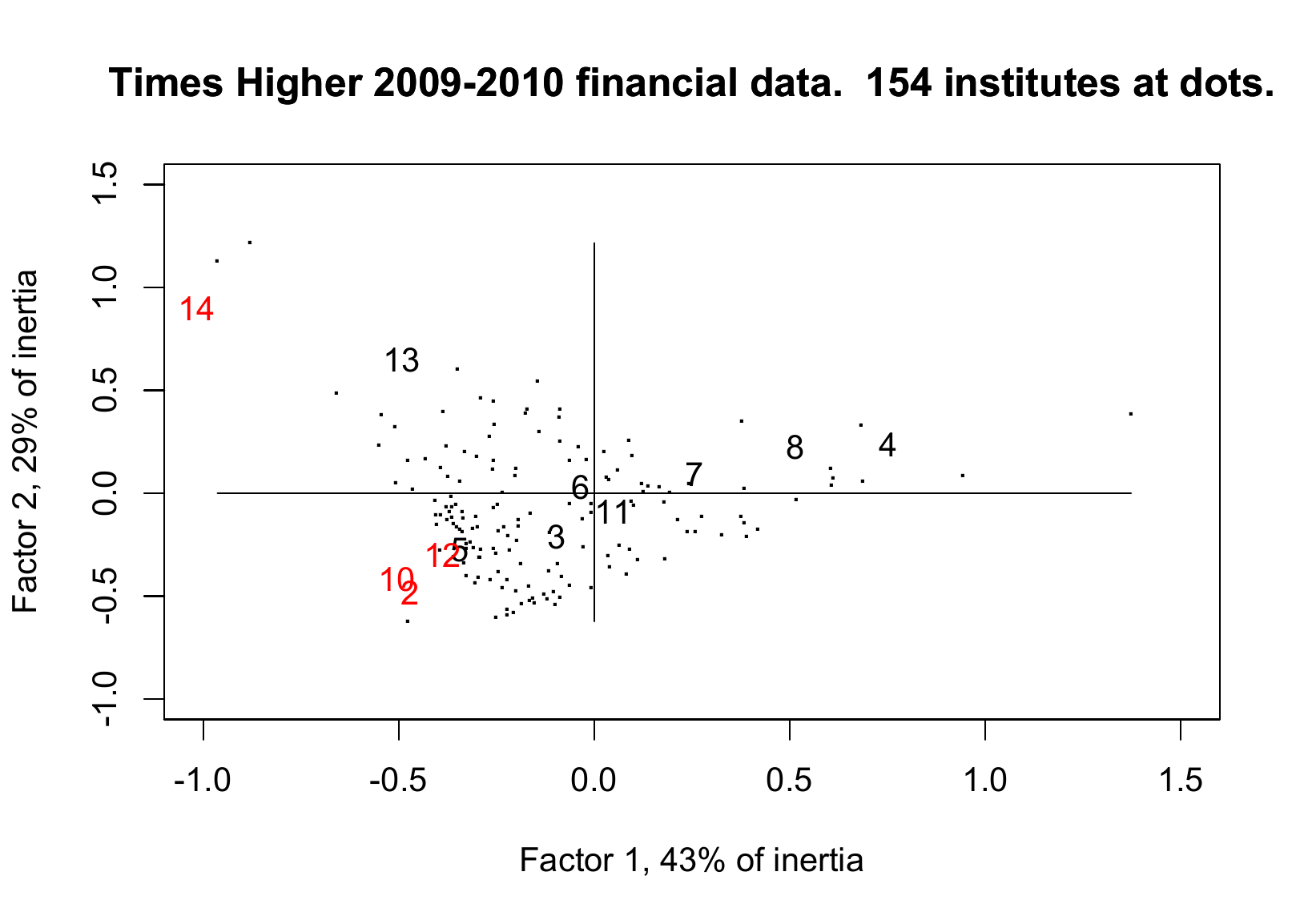}
\caption{Fig.\ \ref{fig1} was related to 2008-2009 and, here, we have 
2009-2010 data.  This is a visualization of the data: the main analysis 
is based on the data table for 154 UK higher education institutions  
using the Times Higher attributes (see text, section \ref{sect2}, 
for these) 
3, 4, 5, 6, 7, 8, 11, and 13.  Their locations can be seen in the
(numeric labels in the) 
planar projection.  In addition, based on the analysis of the 
main data, the locations were found for the more ``illustrative'' 
attributes, 2, 10, 12, and 14 (see also section \ref{sect2}).  
These latter are shown in red.  
The higher education institutes, in order not to crowd this 
initial display, are each shown as a dot.}
\label{fig1new}
\end{figure}

In Baker (2011), data is presented for 2011.  Some (small number of) 
universities differ in the list of 154 used in 2009-2010, compared
to the list of 155 used in 2008-2009.  It is seen though that the
overall characteristics of the data are very similar: cf.\ 
Figs.\ \ref{fig1} and \ref{fig1new}.  

In regard to the rightmost projections on Factor 1 of Fig.\ \ref{fig1new}, 
we again find the following 
(in order of prominence, given by projections).  (Fig.\ \ref{fig3}
had zoomed in on this part of display for the 2008-2009 data.)   

\begin{itemize}
\item Liverpool Sch Tropical Medicine
\item Institute of Cancer Research,
\item Scottish Agricultural College
\item London Sch Hygiene \& Tropical Medicine
\item Oxford
\item Cambridge
\item University of Wales
\item University College London
\end{itemize}

With reference again to the 2009-2010 data, we find the most
prominent on Factor 2 (cf.\ for 2008-2009, Fig.\ \ref{fig4}) to be:

\begin{itemize}
\item Queen Margaret
\item Ravensbourne
\end{itemize}

These are then followed by: Surrey, St Andrews, Worcester,
Reading, Bath, University of London, Bristol.

In regard, for 2009-2010, to the lower left quadrant of 
Fig.\ \ref{fig1new}, and with reference to the year earlier
of 2008-2009 shown in Fig.\ \ref{fig5}, on this occasion 
-- 2009-2010 -- we do not have data for the 
Conservatoire for Dance and Drama (labeled ``CDD'' in Fig.\
\ref{fig4}).   

We do find others though, in order of prominence by projection 
on Factor 1: 

\begin{itemize}
\item London Business School
\item Bath Spa
\item Newman University College
\item Swansea Metropolitan
\item Bishop Grosseteste
\item Manchester Metropolitan
\item Liverpool Inst Performing Arts
\end{itemize}

Overall we see that there is little relative difference between the
two sets of data, for 2008-2009 and 2009-2010.

\section{Implications and Conclusions from the Correspondence Analysis}

We conclude that:

\begin{itemize}
\item  Factor 1 is primarily based on research funding, not from 
HEFCE and sister organizations outside England but rather from 
research councils, and also is  
indicative of the particular importance of medical and bioscience 
research funding which results in institutes that we have noted being 
strongly positioned on this underlying dimension in the data.
As a part of this finding, we note 
this central role played by medical and closely related disciplines.

\item Factor 2 is primarily borrowing, with the
property of gearing (i.e., borrowing relative to income) being
particularly useful to explain this.  Newer institutes,
with limited but focused course offerings, and with specialist
business or industrial sector orientations, together with 
the London Business School, the Open University, and the 
UHI Millennium Institute -- latter now the University of the 
Highlands and Islands -- are all the most extreme in the 
low (or zero) borrowing sense.  
In section \ref{sect4} we have noted the highly geared
institutions.  

\item Our main finding therefore is the polarity between, on 
the one hand, traditional research, by now well swayed 
towards medical and closely related research; and, on the other hand, 
newer and more specialist, or business-oriented institutions.  

\end{itemize}

\begin{figure}
\includegraphics[width=12cm]{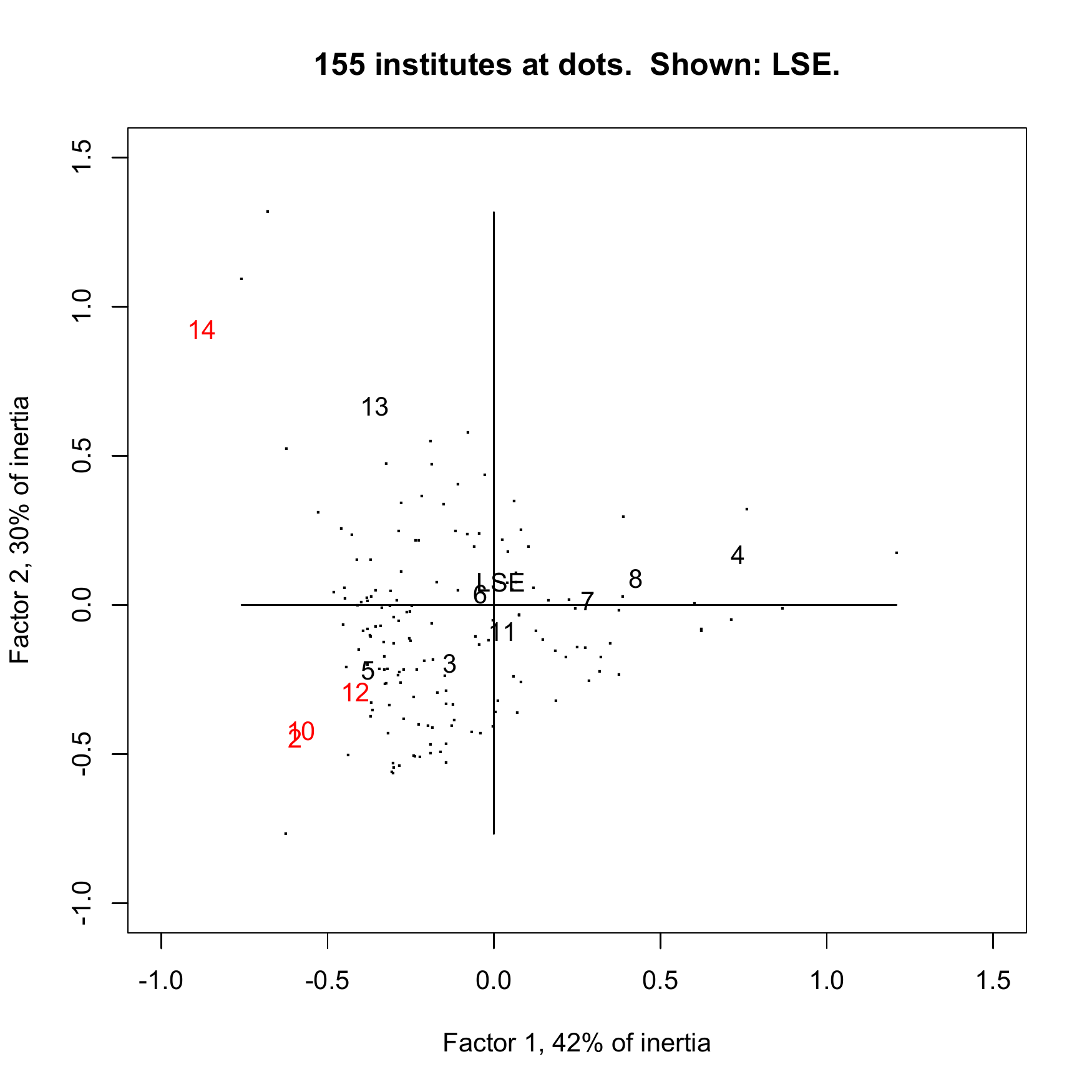}
\caption{Again Fig.\ \ref{fig1}, with LSE (London School of 
Economics) displayed.}
\label{fig5}
\end{figure}

To draw out implications of this polarity we can show -- 
see Fig.\ \ref{fig5} --
the placements of any of the HEIs.
Properties vis-\`a-vis the Factor 1 and Factor 
2 oppositions can be appreciated.  For example, LSE is seen to be in an 
average position.

\section{Further Analyses of Correspondence Analysis Factors}

From the 2009-2010 data, factors 3 and 4 (F3, F4) have this tale to tell.  

\begin{itemize}

\item Attribute 6 at the negative end of F4 relates to Overseas fees.

Attributes 10, 14, then followed by attributes 2, 12 at the positive 
end of F4.  These are all supplementary attributes, i.e.\ projected 
into the analysis passively.  Attributes 2, 12 relate to 
Net surplus as 

\item Attribute 4 somewhat towards the positive end of F3.  This attribute is 
Research grants and contracts other than HEFCE.

Attribute 7 at negative end of F3: Other income, -- catering, 
residential, possibly companies.

F3 therefore distinguishes the sources of income.
\end{itemize}

In summary, F3 is attribute 4 versus attribute 7;  F4 is 
attribute 6 versus all other attributes.  
F4 deals with the 
important, financially sustaining role, of non-European student fee
income.  
F3, as noted, deals with other main sources of income.   

We looked at some further factors and, while interesting for furthering
the study of particular issues, we will not further pursue this here.   

\section{Model-Based Maximum Likelihood Clustering Analysis to 
Specify Clusters of Universities}

\begin{figure}
\includegraphics[width=12cm]{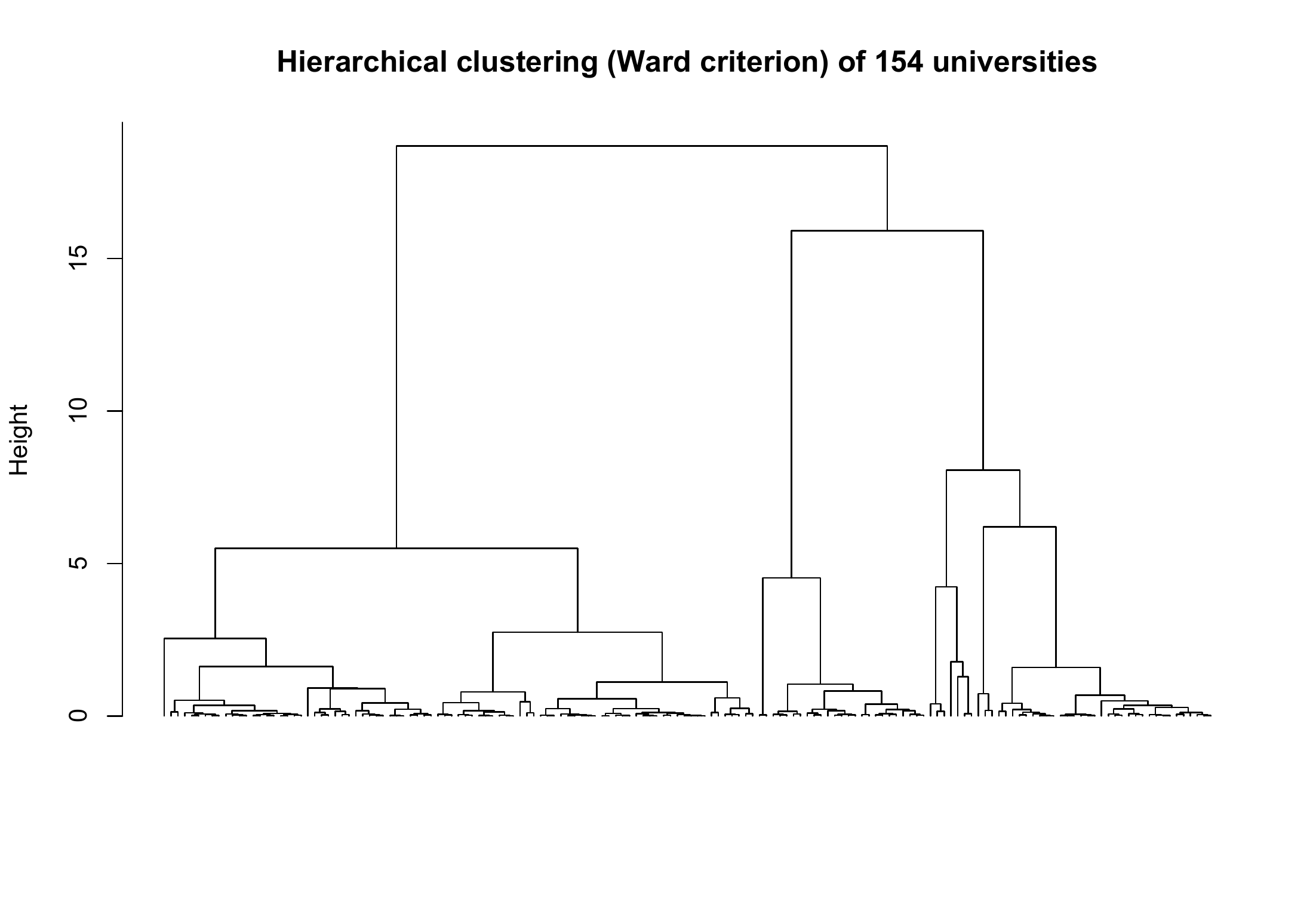}
\caption{Hierarchical clustering, using Ward's minimum variance 
agglomerative criterion, using the full dimensionality, Euclidean
and equi-weighted set of 154 universities, as given by the 
Correspondence Analysis.   2009-2010 data used.}
\label{fig6}
\end{figure}

In Figure \ref{fig6} we use the Correspondence Analysis, and hence 
Euclidean equiweighted, data as input to the clustering.  Furthermore
we use the full dimensionality so there is no loss of information.  

\begin{figure}
\includegraphics[width=12cm]{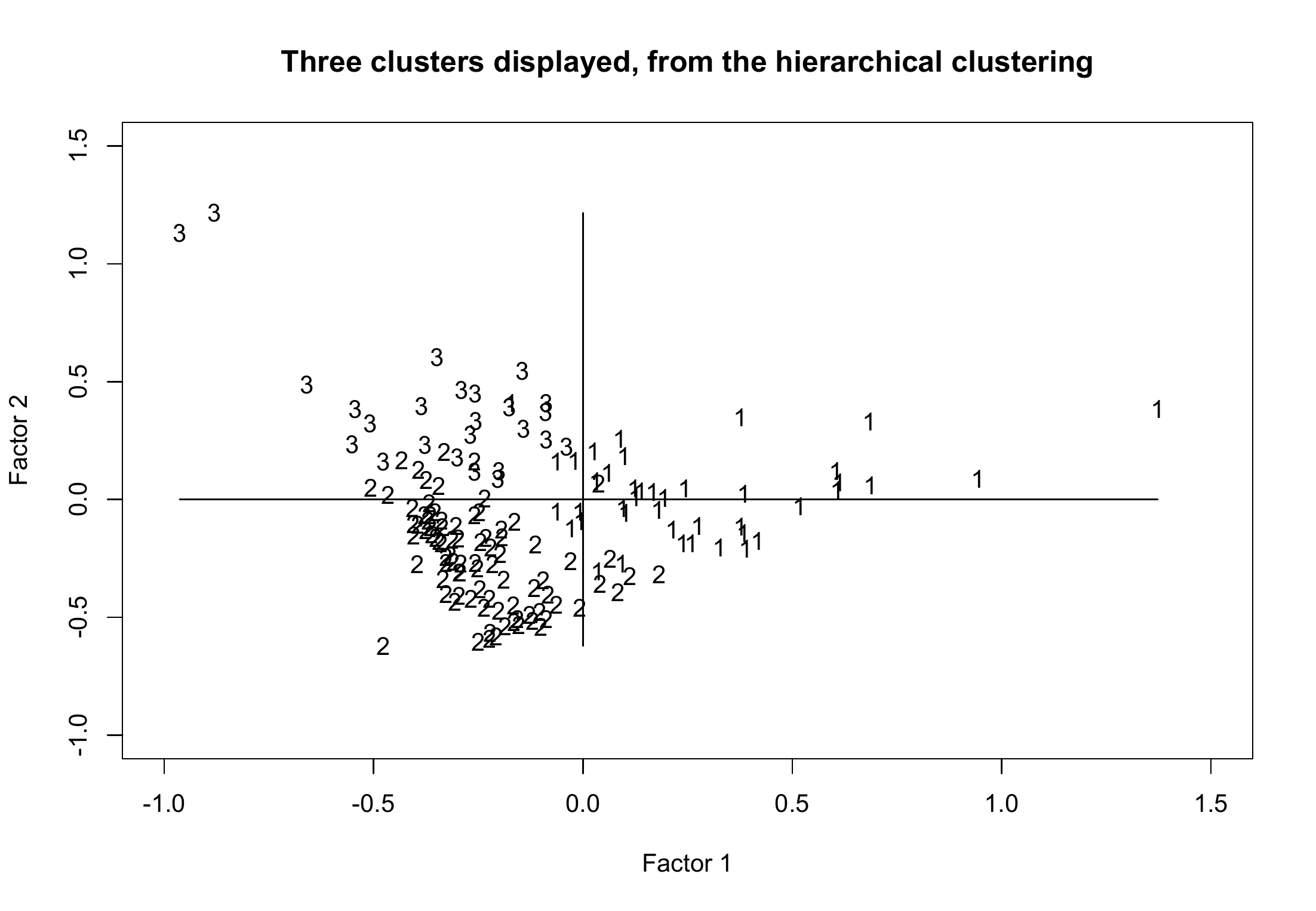}
\caption{Display of the partition with three clusters, derived 
from Figure \ref{fig6}.   The university locations are labeled 
1, 2, and 3, relative to the three clusters.}
\label{fig7}
\end{figure}

The clear three-cluster partition is displayed in the principal 
factor plane in Figure \ref{fig7}.  
As a planar display of the data, this display very much supports
our previous discussion above which was in terms of interpretation
of the factors.  

In order to further support this cluster analysis outcome
we sought corroborating evidence from a Gaussian mixture modeling
of this data.  (We used this approach of hierarchical clustering 
for initial analysis, followed by a model-based approach, in 
Mukherjee et al., 1998). 

The modeling is carried out as follows.  See Fraley and Raftery (1998, 2009).
Take the covariance
matrices, $\Sigma_k$, for cluster $k$.  The eigendecomposition gives 
the decomposition: $\Sigma_k = \lambda_k D_k A_k D_k^t$.  

We use the model that is termed EII, and explained as follows: 

\begin{itemize}
\item Equal volumes for the clusters.
\item Equal shapes.  
\item Orientation of clusters is not relevant due to sphericity. 
\end{itemize}

The model in this case is $\Sigma_k = \lambda I$ where $I$ is the 
identity matrix, and $\lambda$ is the same eigenvalue for all 
clusters $k$.  Informally we are fitting hyperspherical balls 
of the same characteristics to our data.  In this way, we 
determine the cluster components that, when aggregated, give rise 
to the complicated cluster morphologies that are observed in practice. 

\begin{figure}
\includegraphics[width=12cm]{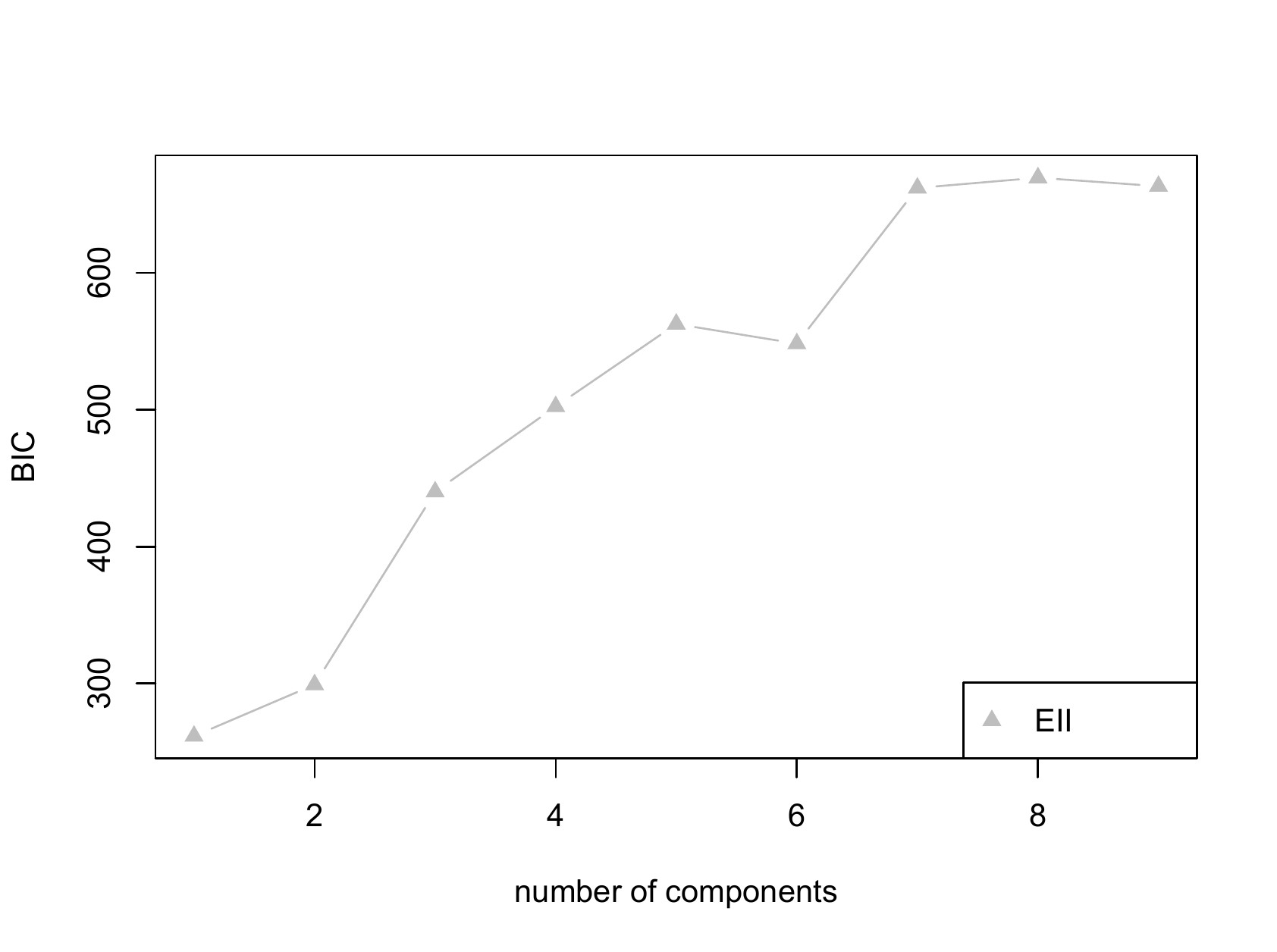}
\caption{Bayesian information criterion, BIC, pointing to a best fit of 
8 clusters for this EII (see text for details) model.}
\label{fig8}
\end{figure}

The Bayesian information criterion, BIC, is used for model identification.   
Figure \ref{fig8} shows the outcome.

\begin{figure}
\includegraphics[width=12cm]{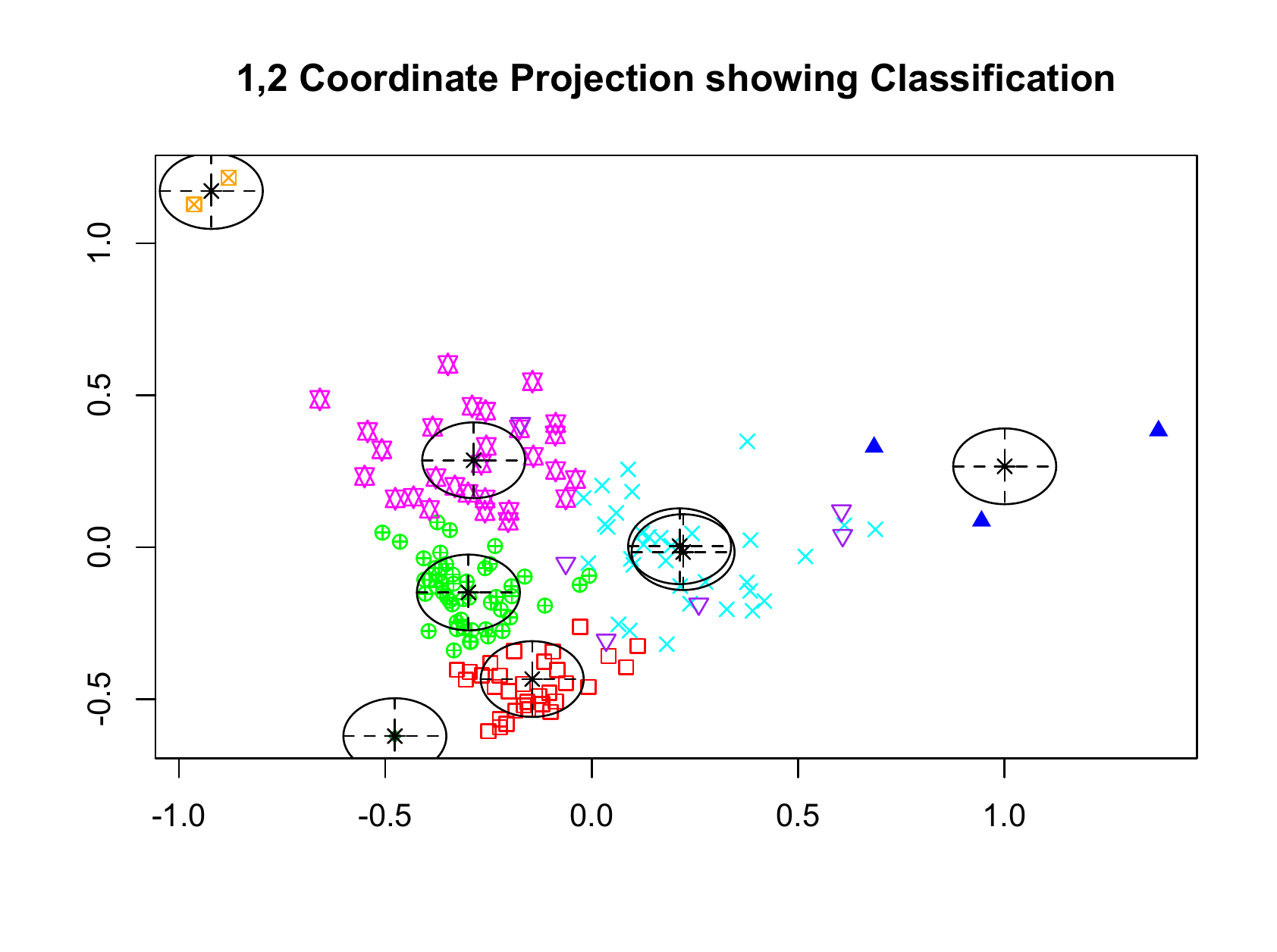}
\caption{A principal coordinate projection of the 8 clusters found
through fitting of the EII cluster model to the data.  The 
clusters here are ``components'' of the complex morphologies that
are found in practice.}   
\label{fig9}
\end{figure}

In Figure \ref{fig9}, we see the clusters projected in a principal 
coordinate plane.  The principal coordinates were determined from the 
Euclidean Correspondence Analysis factor output, so the same outcome 
as we have in earlier figures is to be expected. 
The support that is found using this approach, relative
to e.g.\ Figure \ref{fig7}, is strong.   
In fact we see how very well these clusters in Figure \ref{fig9} 
related to the displays of Figures \ref{fig2} (right hand side), 
\ref{fig3} (upper left hand side) and \ref{fig4} (lower left hand side).


\section{Discussion and Conclusions}

In studying world leadership in research, Adams and Gurney (2010) 
find five institutes (Oxford, Cambridge, UCL, Imperial, and LSE)
to be significantly separate from all others, including others in
the Russell Group of universities.  Adams and Gurney label the 
five universities the ``Golden Triangle''.  The criterion used 
by Adams and Gurney is citation impact, based on Thomson Reuters
databases.   Of course this is not necessarily a good basis for 
the measurement of impact in, for example, computer science (see e.g.\ 
Moed and Visser, 2007) due to more limited coverage of the literature 
in this area and also different citation practices and culture (involving 
books and conferences, for example), and other disciplines can be 
added, in engineering, mathematics and the humanities.  

When viewing the university system in its entirety, other 
forms of impact are clearly important also.  These include 
human capital, sectoral and niche applications, and also 
engineering (as opposed to science) demonstrators and
testbeds, and their deployment.  

It is seen from our data analysis that the UK system is 
gravitating -- in fact, it has largely already done so -- towards two 
attractors: high research income, and what we have characterized as 
niche industrial/business
sector application-oriented research, that also incorporates
business and management, and human capital too.
These corresponding to the right hand side of our displays, and 
to the lower left hand side, respectively.   

Note that all of these
planar projection displays are not invariant from the interpretation
point of view, relative to a reflection symmetry about the axes.   
However for a given software implementation they are of course 
replicable. 

Let us take one step further 
our findings in regard to these university ``attractors''.  We 
raise the question of what are appropriate performance metrics.
On the one hand, impact of funded research, as measured through citations, 
which as a performance measurement tool is very fit for purpose 
across a wide range of disciplines 
including the life sciences, biosciences, materials science and 
others.  On the other hand, performance that is evaluated by a 
narrative of impact is what is coming about in regard to outputs and 
outcomes from what we have characterized as niche industrial/business
sector application-oriented research, that also incorporates 
business and management, and human capital.

\section*{References}

\begin{enumerate}

\item 
Jonathan Adams and Karen Gurney (2010).  
``Funding selectivity, concentration and excellence -- how good is the UK's 
research?'', 
{\em HEPI report, Higher Education Policy Institute}, 20 pp., 
25 March 2010.

\item Simon Baker (2011). 
``Underlying disorders: The financial health of UK higher 
education institutions'',  {\em Times Higher Education}, 7 April 2011.  
Data tables in ``Wealth and Health: Financial Data for UK Higher Education
Institutions, 2009-2010''.  

\item Ronald Barnett (2003). Beyond All Reason: Living with Ideology in the University, 
Society for Research into Higher Education and Open University Press. 

\item John Beath, Joanna Poyago-Theotoky and David Ulph (2011). 
``University funding systems: 
impact on research and teaching'', Discussion Paper Nr. 2011-1, 
http://www.economics-ejournal.org/economics/discussionpapers/2011-1, 
submitted to {\em Quasi Markets in Education}.  

\item 
BIS, Department for Business Innovation \& Skills (2010). 
The Allocation of Science and 
Research Funding 2011/12 to 2014/15, Investing in World-Class Science and Research, 
Dec. 2010, 59 pp. 

\item Chris Fraley and Adrian E. Raftery (1998). 
``How many clusters? Which clustering methods? Answers via 
model-based cluster analysis'', {\em Computer Journal}, 41, 578--588. 

\item Chris Fraley and Adrian E. Raftery (2009), 
``MCLUST Version 3 for R: normal mixture
modeling and model-based clustering'', Technical Report No.\ 504, Department
of Statistics, University of Washington, December 2009 version, pp. 56. 

\item Sachi Hatakenaka (2005). ``Development of third stream activity.   Lessons from 
international experience'',  
{\em HEPI report, Higher Education Policy Institute}, Nov. 2005.

\item Alan Jenkins (2004). A Guide to the Research Evidence on Teaching-Research 
Relations, The Higher Education Academy, York, UK, Dec. 2004, 36 pp.  

\item Alan Jenkins and Mick Healey (2005). 
Institutional Strategies to Link Teaching and 
Research, The Higher Education Academy, York, UK, 66 pp.  

\item F. Lebaron (2011).``Geometric data analysis in a social science research program: The    
case of Bourdieu's sociology'', in M. Gettler Summa, L. Bottou, B. Goldfarb, F. Murtagh, 
C. Pardoux and M. Touati, {\em Statistical Learning and Data Science}, Chapman \& Hall/CRC 
Press, 2011.  pp. 77--89.

\item Trevor McMillan, Tom Norton, Justin B. Jacobs and Rosanagh Ker
(2010), on behalf of the 
1994 Group's Research \& Enterprise Policy Group, Enterprising Universities, Using the 
Research Base to Add Value to Business, Policy Report, Sept. 2010, 1994 Group, London, 22 pp. 

\item Henk F.\ Moed and Martijn S.\ Visser (2007). ``Developing bibliometric 
indicators of research performance in computer science: an exploratory 
study'', Research report to the Council for Physical Sciences of the
Netherlands Organisation for Scientific Research (NWO), 117 pp., Feb.\ 2007.

\item 
Soma Mukherjee, Eric D. Feigelson, Gutti Jogesh
Babu, Fionn Murtagh, Chris Fraley and Adrian
Raftery (1998), 
``Three types of gamma ray bursts'', 
{\em The Astrophysical Journal}, 508, 314--327, 1998. 

\item 
Fionn Murtagh (2010). ``The Correspondence Analysis platform for 
uncovering deep structure in data and information'', 
Sixth Annual Boole Lecture in Informatics,  
{\em Computer Journal}, {\bf 53} (3), 304--315.  

\item Melanie Newman (2010). ``Ready for the storm?'', {\em 
Times Higher Education}, 18 
March 2010, pp.\ 36-44.

\end{enumerate}

\end{document}